\documentstyle[aps,prb,epsf,multicol]{revtex}
\begin{document}

\newcommand{\hommet}  {HM}
\newcommand{\hommetsg}{HM$_{\text{sg}}$}
\newcommand{\homins}  {HI}
\newcommand{\hominssg}{HI$_{\text{sg}}$}
\newcommand{\cdw}     {CDW}
\newcommand{\cdwsg}   {CDW$_{\text{sg}}$}

\title{
Phase diagram of the quarter--filled extended Hubbard model
on a two--leg ladder
}
\author{Matthias Vojta$^{(a)}$\cite{addr}, Arnd H\"ubsch$^{(b)}$, and
R.~M. Noack$^{(c)}$}
\address{
  (a) Department of Physics, Yale University,
      New Haven, CT 06520--8120, USA \\
  (b) Institut f\"{u}r Theoretische Physik,
      Technische Universit\"{a}t Dresden, D--01062 Dresden, Germany \\
  (c) Institut f\"ur Physik, Johannes--Gutenberg--Universit\"at Mainz,
      D--55099 Mainz, Germany
}
\date{\today}
\maketitle

\begin{abstract}
We investigate the ground--state phase diagram of the quarter--filled
Hubbard ladder with nearest--neighbor
Coulomb repulsion $V$ using the Density Matrix
Renormalization Group technique.
The ground--state is homogeneous at small $V$,
a ``checkerboard'' charge--ordered insulator at large $V$ and not
too small on--site Coulomb repulsion $U$,
and is phase--separated for moderate or large $V$ and small $U$.
The zero--temperature transition between the homogeneous and the
charge--ordered phase is found to be second order.
In both the homogeneous and the charge--ordered phases
the existence of a spin gap mainly depends on the
ratio of interchain to intrachain hopping.
In the second part of the paper,
we construct an effective Hamiltonian for the spin degrees of freedom
in the strong--coupling charge--ordered regime which maps the system
onto a frustrated spin chain.
The opening of a spin gap is thus connected with spontaneous dimerization.
\end{abstract}

\pacs{PACS: 71.10.Fd, 71.10.Pm, 71.27.+a}

\begin{multicols}{2}
\narrowtext
\newpage


\section{Introduction}

Quasi--one--dimensional systems present a unique opportunity to study
the interplay between strong quantum fluctuations on the one hand and
a tendency to charge or spin ordering on the other.
Examples include Peierls or spin--Peierls behavior as well as
charge ordering due to electron--electron interaction.
One material in which such behavior has been found is the two--leg
ladder material NaV$_2$O$_5$.
NaV$_2$O$_5$ undergoes a phase transition at $T_c=34$ K that is characterized
by the opening of a spin gap and a doubling of the unit cell.
Although this transition was originally thought to be spin--Peierls,
recent studies have found evidence for charge
order.\cite{ohama,isobe,neutron}
Above the transition, the material seems to be best described
as a quarter--filled ladder.\cite{Smol98,Seo98}
Such two--leg--ladder structures have also been found in
a number of other materials, including the
Vanadates MgV$_2$O$_5$ and CaV$_2$O$_5$, and the
cuprates SrCu$_2$O$_3$ and Sr$_{14}$Cu$_{24}$O$_{41}$.
For a more detailed description of ladder materials and models as
well as a discussion of the extensive theoretical work on ladder
models,
we direct the reader to a recent review \cite{dagottoreview}
and the references contained therein.

In NaV$_2$O$_5$, the character of the charge ordering and the nature of
the transition are currently under debate.
While evidence is growing \cite{ohama2,nakao,grenier00} that the
charge--ordered
state has a zigzag--like charge arrangement, the question of whether this
charge ordering is continuous or discontinuous has not yet been settled
experimentally.
Theoretically, the magnetic properties of the compound for $T<T_c$
have been calculated for different charge--ordering patterns
\cite{gros,thal2} and compared to experiments.
Again, zigzag charge ordering on ladders (possibly with intervening
disordered ladders \cite{smaalen}) leads to magnon dispersions
in good agreement with neutron scattering data.
Additionally, the possibility of two separate transitions at nearby
temperatures has been raised;\cite{steglich1,thal1} in this
scenario the charge order would set in at one temperature while the
spin gap would open at a second, lower critical $T$.
Very recently, a scenario of singlet cluster formation instead
of zigzag charge ordering has been suggested. \cite{boer00}
However, this model seems not to be able to reproduce the experimentally
found spin gap and magnon dispersion data. \cite{gros00}
It is clear that more experimental work is necessary to clarify the
nature of the low--temperature phase of NaV$_2$O$_5$.

One of the simplest models of interacting electrons that allows for
charge ordering is the extended Hubbard model, i.e.\ the Hubbard model
supplemented by an additional nearest--neighbor (NN) Coulomb
repulsion, $V$.
This model has been studied in one dimension (1D) in
the strong--coupling limit,\cite{hubbard78} at quarter filling
\cite{Mila93,Penc94} and at half filling
\cite{Cannon91,hirsch,Zhang97,tricrit}
and in between,\cite{Lin95,Clay99}
in the 2D system at half filling,
\cite{Zhang89,Chatto97} and within the Dynamical Mean Field Theory (the
limit of infinite dimensions) at quarter \cite{Pietig99}
and half filling. \cite{Dongen94}
A variety of techniques, such as mean--field approximations,
perturbation theory, as well as numerical methods as quantum Monte
Carlo and the Density Matrix Renormalization Group (DMRG) have been
employed.

The result of investigations of the charge--order transition can be
summarized as follows:
At the mean--field level, the transition between a homogeneous
state and a charge--density wave (CDW) state at half filling in a
hypercubic lattice occurs at $V_c = U/z_0$, where
$z_0$ denotes the number of nearest neighbors ($z_0 = 2d$) and $U$ is
the on--site interaction.
Numerical studies \cite{hirsch,Zhang97} indicate a slightly higher value of
$V_c$, at least in 1D.
Interestingly, the transition at half filling in 1D has been found to
be second order at small $U/t$ and first order at large $U/t$ with the
tricritical point located at $U_c/t \sim 4-6$. \cite{Zhang97,tricrit}
Here we use the term ``first order'' to denote discontinuous behavior
of the charge order parameter as a function of microscopic
parameters such as $V$ or band filling, and ``second order'' to denote
continuous behavior.
For fillings below half--filling in 1D, the situation is more
complicated because a number of phases compete at large
$V$. \cite{Mila93,Penc94,Clay99}
For dimension larger than one, indications are that the charge--order
transition is generally first order.
\cite{Zhang89,Chatto97,Pietig99,Dongen94}
However, conclusive studies that can
reliably distinguish between first-- and second--order transitions are
lacking.
At small $U$ and large $V$, the extended Hubbard model undergoes phase
separation (PS) rather than a transition to a CDW state.
For the 1D model between quarter-- and half--filling,
it has been established\cite{Clay99} that PS occurs for $|U|/t < 4$
in the $V=\infty$ limit, whereas for $U/t > 4$ the system undergoes a
transition to a $q=\pi$ CDW state
for sufficiently large $V$.\cite{Lin95}
Phase separation in higher dimensions has also been
discussed. \cite{Dongen96}

For the 1D system in particular there has been recent interest in
the possibility of dominant superconducting correlations in the
uniform ground--state away from half--filling when
$V \gg U \sim t$, \cite{Penc94,Clay99}
i.e., in the proximity of the phase--separated region.
We note here that the uniform phase in 1D off half--filling is metallic
and can in general be described within the Luttinger--liquid picture.
Although dominant superconducting correlations
have not been established in the ground state of the
1D extended Hubbard model to date, a number of non--Luttinger--liquid effects
have been observed.\cite{Clay99}

Some of the present authors have previously studied \cite{VHN99} the
charge--order transition in the extended Hubbard model on the two--leg
ladder at various band fillings for $U/t=4$ and 8.
A transition to a checkerboard charge--ordered state was found for
all fillings between quarter-- and half--filling.
The transition is second--order near quarter filling and
first--order near half--filling for sufficiently large $U$.

The focus of the present paper is on this model at quarter filling,
with a two--fold purpose:
First, we present a comprehensive study of the phase diagram as a
function of $U/t$ and $V/t$ for repulsive $U$ and $V$ and discuss
the properties of the ground state phases as well as the nature
of the phase transitions.
Second, we derive an effective Hamiltonian for the spin degrees
of freedom in the charge ordered state at strong coupling, and
compare the predictions of this effective low--energy theory with
our numerical results.

\subsection{Phase diagram}

\begin{figure}
\epsfxsize=7.5 truecm
\centerline{\epsffile{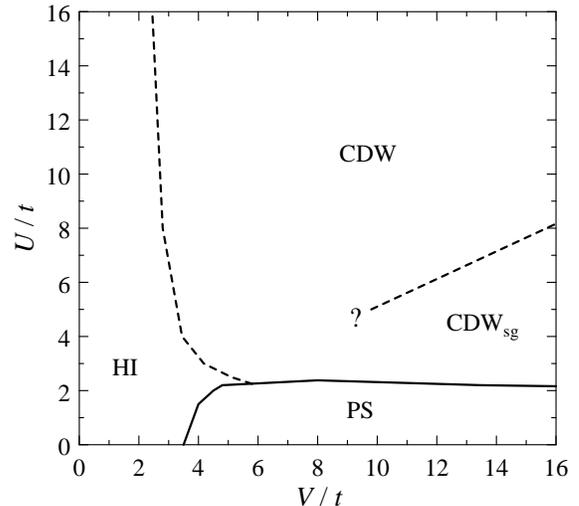}}
\caption{
  Ground state phase diagram of the extended quarter--filled
  Hubbard model on a two--leg ladder with isotropic hopping,
  as a function of the on--site and nearest--neighbor
  repulsion, $U/t$ and $V/t$.
  The phase labeling is explained in the text,
  the dashed lines represent second order
  phase transitions.
  The solid line marks the boundary of the phase separation region (PS)
  where the thermodynamic compressibility diverges.
}
\label{figpd}
\end{figure}

\begin{figure}
\epsfxsize=8.8 truecm
\centerline{\epsffile{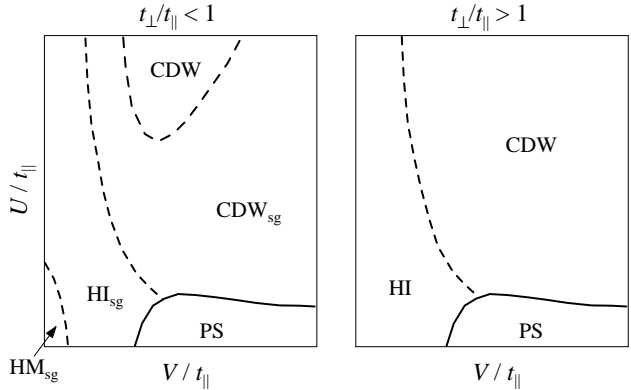}}
\caption{
  Proposed schematic phase diagrams for the extended quarter--filled
  Hubbard model with (large) hopping anisotropy.
  The phases are labeled as before.
  For $t_\perp < t_\parallel$, the spin gap is
  nonzero in accordance with the weak coupling predictions,
  whereas for $t_\perp > t_\parallel$ the spin gap is always zero.
  In the former case, the spin gap can be suppressed deep
  in the charge--ordered phase if $V/U$ is smaller than
  a critical value.
}
\label{figpdtr}
\end{figure}

Our main results, the phase diagrams deduced from the numerical
calculations, are summarized in Figs.~\ref{figpd} (isotropic hopping)
and \ref{figpdtr} (anisotropic hopping).
The phases are distinguished by the presence or absence of a gap
for spin and/or charge excitations.
To denote this, we employ the following labeling:
\hominssg\ denotes a homogeneous insulator (nonzero charge gap)
with a spin gap, \homins\ a homogeneous insulator without a spingap,
\hommetsg\  (\hommet) a homogeneous metallic phase having zero charge gap
with (without) spin gap, and
\cdwsg\ (\cdw) is a charge--ordered state with (without) spin gap.
The CDW states are always insulating in the present quarter--filled
model.
The phase diagrams can be roughly divided into four regions:
(i) Weak coupling: for small $U$ and $V$ we find homogeneous phases
similar to the ones in the ``bare'' Hubbard model (see discussion
in Sec.~\ref{sec:model} and results in Sec.~\ref{sec:hom}).
(ii) Large $U$, small $V$: These homogeneous strong coupling phases
have characteristics similar to the weak--coupling region.
(iii) Small $U$, large $V$: phase separation, this is discussed further
in Sec.~\ref{sec:ps}.
(iv) Strong coupling: large $U$ and $V$ lead to an insulating
checkerboard charge--ordered with either gapless or gapped
spin excitations depending on the ratio of $V/U$.

For isotropic hopping we have numerically determined the phase
boundaries as shown in Fig.~\ref{figpd}.
The precise location of the \cdw--\cdwsg\ boundary at which the spin gap
closes at intermediate coupling could not be obtained by the
methods used here; we have indicated this uncertainty
by a question mark in the phase diagram.

In the case of anisotropic hopping we have not mapped out the
full phase diagram, but the numerical results (discussed in
Sec.~\ref{sec:results}) provide the schematic pictures
shown in Fig.~\ref{figpdtr}.
Varying the ratio of the rung to leg hopping strengths,
$t_\perp/t_\parallel$, has two effects:
(a) For small $t_\perp/ t_\parallel$ there appears a metallic
phase (\hommetsg) with spin gap and dominating d--wave--like singlet pair
correlations (as in the ``bare'' $V=0$ Hubbard ladder).
(b) The existence of a spin gap depends strongly on
the hopping ratio, i.e., there is a transition
as function of $t_\perp / t_\parallel$ where a spin gap
opens (in both the homogeneous and CDW phases).
The critical hopping ratio may depend on the interaction
strength, but is near unity in the homogeneous phases for
reasonable values of the interactions.

The rest of the paper is organized as follows:
In Sec.~\ref{sec:model}, we introduce the extended Hubbard model
and discuss some results known for the $V=0$ case, i.e., the ``bare''
Hubbard model on a two--leg ladder.
In Sec.~\ref{sec:results}, we present our numerical results,
discuss the properties of the phases shown in Figs.~\ref{figpd}
and \ref{figpdtr},
and examine the transition to the charge--ordered state.
At large $V$, where charge order is well established, it is possible
to derive an effective Hamiltonian for the residual spin degrees
of freedom; this is done in Sec.~\ref{sec:strong}.
A summary and a discussion of the relevance of our results to
experimental systems (especially NaV$_2$O$_5$)
terminates the paper.


\section{Extended Hubbard model}
\label{sec:model}

The single--band extended Hubbard model has the Hamiltonian
\begin{eqnarray}
H&=&- \sum_{\langle ij\rangle\sigma} t_{ij}
(c^\dagger_{i\sigma} c_{j\sigma}^{} + h.c.)
\nonumber\\
 && + U\sum_{i} n_{i\uparrow} n_{i\downarrow}
    + \sum_{\langle ij\rangle} V_{ij} n_i n_j \; .
\label{hamiltonian}
\end{eqnarray}
Here we consider a lattice consisting of two chains of length $L$,
i.e., a ladder, and restrict ourselves to the band filling
$\langle n \rangle = N/(2L) = 1/2$,
where $N$ is the number of electrons.
The summation $\langle ij\rangle$ runs over all pairs of nearest--neighbor
sites on the ladder, taking open boundary conditions between the chains.
The hopping matrix elements along the legs and rungs of the ladder are
denoted $t_\parallel$ and $t_\perp$, respectively, and
the nearest--neighbor Coulomb interactions are similarly denoted
$V_\parallel$ and $V_\perp$.
Unless otherwise noted, we will use $t_\parallel$ as unit of energy.
In this work, we will treat primarily the ``isotropic'' case,
$t_\parallel = t_\perp = t$ and $V_\parallel = V_\perp = V$.

The non--interacting Hamiltonian ($U=V=0$) can be diagonalized by
a Fourier transform (for periodic boundary conditions in the chain
direction), leading to the single--particle energies
\begin{equation}
\epsilon_{\bf q} = -2t_\parallel \cos q_x + t_\perp \cos q_y
\end{equation}
where ${\bf q}=(q_x,q_y)$, $q_x$ is the momentum along the chains
and the momenta $q_y = 0,\,\pi$ correspond to bonding and
antibonding symmetry, respectively.
Either one or both of the bands can be occupied in the noninteracting
system, depending on the total particle density and the ratio of $t_\perp$
and $t_\parallel$.
At quarter filling, the transition occurs at
isotropic hopping:
for $t_\perp < t_\parallel$ both bands are less than half--filled, whereas
for $t_\perp > t_\parallel$ the bonding band is half--filled
and the antibonding band is unoccupied.

The effect of the Hubbard interaction $U$ on this system has been
extensively studied.
In the weak--coupling limit, $U\ll t$, the phase diagram has been
investigated using the perturbative renormalization
group (RG). \cite{bfrg}
A variety of phases have been shown to exist
as a function of band filling and hopping anisotropy.
In general, the two--band system can have four possible modes
(symmetric and antisymmetric charge and spin modes), each of which can
be either massive or massless.
The phases can therefore be classified using the notation C$n$S$m$
where $n$ and $m$ designate the
number of gapless charge and spin modes, respectively ($0\le n,m \le 2$).
At quarter filling, the weak--coupling RG \cite{bfrg}
for the ``bare'' Hubbard model ($V=0$) yields the following results:
For $t_\perp > t_\parallel$ the system behaves as
a half--filled Luttinger liquid; Umklapp scattering in the bonding
channel is a relevant perturbation which leads to a charge--gapped C0S1
phase at small $U$.
In contrast, deep in the two--band region, $t_\perp \ll t_\parallel$,
one finds a metallic C1S0 phase in the weak--coupling limit.
Near isotropic hopping, the bottom of the antibonding band just ``touches''
the Fermi surface, and the curvature of the dispersion becomes
important, leading to additional narrow regions of C2S2 and C2S1 phases.
Several of these weak--coupling predictions have been verified
by numerical DMRG calculations in the intermediate and strong
coupling regimes for a wide range of filling. \cite{noacktwochains}
Systematic studies of the phases of the {\em extended} Hubbard model
on a two--leg ladder away from half--filling have to our knowledge
not yet been carried out.


\section{Numerical results}
\label{sec:results}

In this section we present the results of our numerical
investigations and discuss the characteristics of the phases
shown in Figs.~\ref{figpd} and \ref{figpdtr}.

\subsection{Technique and observables}

The numerical results have been calculated with
the DMRG technique \cite{DMRG} on lattices of up to $2\times 80$
sites with open boundary conditions at
the ends of the chains as well as between the two chains.
Most data shown are obtained by keeping 600 states per block,
resulting in the sum of the discarded density matrix eigenvalues
being typically $10^{-8}$ or less.
For small system sizes we have checked the convergence by using
up to 1000 states per block.
Unless otherwise noted, we estimate the errors in the gap energies and
correlation functions obtained using the DMRG procedure to be less than a
few percent.

Important ground--state properties are the static charge and
spin correlation functions: we have calculated the
static charge structure factor
\begin{eqnarray}
C({\bf q}) = {1 \over 2 L} \sum_{i}
  e^{i {\bf q} \cdot {\bf R}_i} \bar{C}({\bf R}_i)
\end{eqnarray}
where
\begin{equation}
\bar{C}({\bf R}_i) = \frac{1}{N_{\rm av}}\sum_{\{j \}}
\langle \delta n_{j+i} \delta n_j\rangle  \; ,
\end{equation}
$\langle ... \rangle$ denotes the ground--state expectation value,
$\delta n_j = n_j - \langle n_j \rangle $, and we average over
typically $N_{\rm av} = 6$ sites to remove oscillations due to the open
boundaries.
The spin structure factor $S({\bf q})$ is defined similarly in terms of
the spin--spin correlation function $\langle S_{j+i}^z S_j^z \rangle$.

The nature of the low--lying excitations can be determined by calculating
the energy gaps of the system.
In particular, we will consider the charge and spin gaps, defined as
\begin{eqnarray}
\Delta_c &=& \frac{1}{2} \left[
E_0(L,N+2)+E_0(L,N-2)-2E_0(L,N)
\right] \,, \nonumber\\
\Delta_s &=& E_0(L,N,S_z=1) - E_0(L,N,S_z=0)
\label{gapdef}
\end{eqnarray}
where $E_0(L,N)$ is the ground state energy of ladder system
with $2L$ sites and $N$ electrons.
Since we calculate the gaps $\Delta(L)$ on finite systems,
the gaps must be extrapolated to $L\rightarrow\infty$;
we do this by performing a polynomial fit in $1/L$ through the data
points from the larger system sizes ($L\ge 24$).
Although we include both $1/L$ and $1/L^2$ terms, the coefficient of
the quadratic term is quite small in most cases.
The uncertainty of the extrapolated value depends strongly on
finite--size effects which become large when the correlation length
becomes large;
the results for the gaps are most accurate in the strong coupling region,
$U,V \gg t$, and for not too small $t_{\perp}$.

\subsection{Homogeneous phases}
\label{sec:hom}

\begin{figure}
\epsfxsize=8 truecm
\centerline{\epsffile{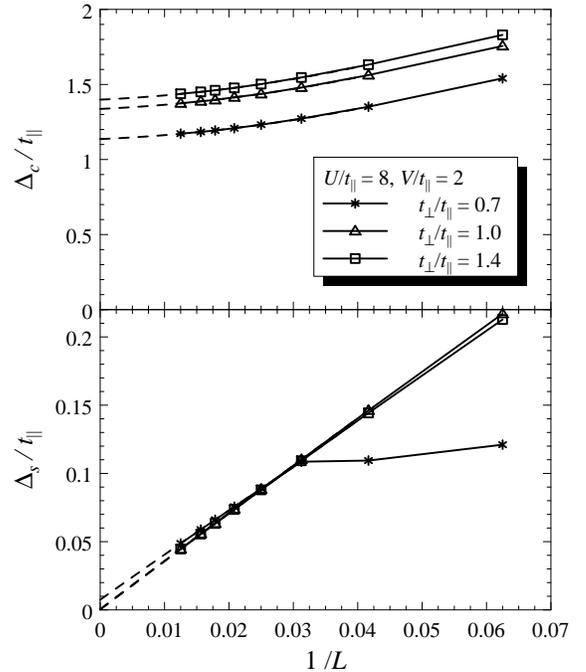}}
\caption{
  Finite--size scaling for the charge and spin gaps
  at $U/t_\parallel=8$, $V/t_\parallel=2$ and different
  $t_\perp/t_\parallel$.
  The solid lines are guides to the eye,
  the dashed lines are quadratic fits through the data points for
  $L=24$ through $80$.
}
\label{figgapscale}
\end{figure}

First, we concentrate on the states without charge order, i.e.,
the region of small $V$ as shown in Fig.~\ref{figpd}.
The calculated charge and spin correlation functions (see e.g.
Fig. 1 of Ref. \onlinecite{VHN99}) indicate antiferromagnetic
correlations peaked at ordering vector $(\pi,\pi)$.
The nature of the phases is best probed by calculating
spin and charge gaps.
The extrapolation to the thermodynamic limit
is illustrated in Fig.~\ref{figgapscale}, in which
we show results for charge and spin gaps at $U/t_\parallel=8$,
$V/t_\parallel=2$ and different values of the hopping ratio
$t_\perp/t_\parallel$.
As noted above, finite--size effects increase with decreasing
interchain coupling $t_\perp$.
However,
we have verified (by keeping more DMRG states per block and/or
using a fit with a $1/L$ term only) that, within the numerical
accuracy available, $\Delta_s$ in Fig.~\ref{figgapscale}
vanishes for $t_\perp/t_\parallel=1$, but
is nonzero for $t_\perp/t_\parallel=0.7$.

\begin{figure}
\epsfxsize=7.5 truecm
\centerline{\epsffile{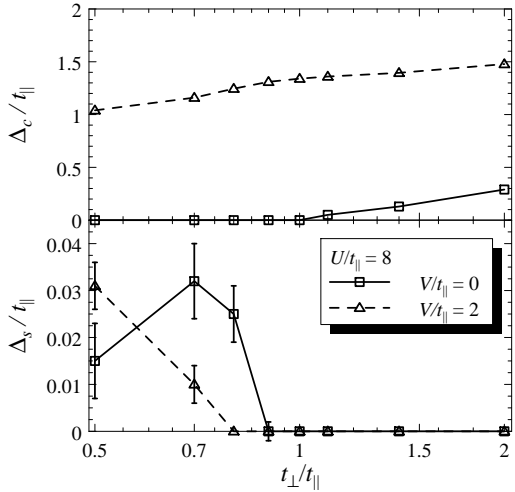}}
\caption{
  Charge and spin gaps at $U/t_\parallel=8$, $V/t_\parallel=0, 2$ as
  function of the hopping ratio $t_\perp/t_\parallel$.
  Since the finite--size effects become substantial for small $t_\perp$,
  we have indicated the estimated errors from the extrapolation to the
  thermodynamic limit by error bars.
  For data points without error bars, the uncertainties are of the
  order of the symbol size or less.
}
\label{figgaptp}
\end{figure}

The extrapolated DMRG results for charge and spin gaps in the homogeneous
phase are displayed in Fig.~\ref{figgaptp}.
First we discuss the $V=0$ case, i.e., the ``bare'' Hubbard ladder.
According to the weak coupling RG, \cite{bfrg}
both spin and charge gaps vanish in the case of isotropic hopping.
Varying the hopping ratio $t_\perp/t_\parallel$ tunes the
system through the one--band to two--band transition:
For $t_\perp > t_\parallel$ Umklapp scattering opens a charge gap
(C0S1 phase); with decreasing $t_\perp/t_\parallel$ one finds
narrow regions of C2S2 and C2S1, followed by a C1S0 phase.
Our results for $U/t=8$ agree with these predictions:
we find a C0S1 (\homins) phase for $t_\perp > t_\parallel$
and a C1S0 (\hommetsg)
phase for $t_\perp < 0.9 t_\parallel$.
The data also indicate a narrow region where both $\Delta_c=0$ and
$\Delta_s=0$ (\hommet), possibly corresponding to the C2S2 and C2S1 phases.
Turning to $V>0$,
the described behavior continues to small nonzero values of $V$, but
the transition points shift to smaller $t_\perp/t_\parallel$.
A further increase of $V$ suppresses the metallic phase, and
only the spin gap transition remains.
Data for $V/t_\parallel=2$ is shown in Fig.~\ref{figgaptp}:
the behavior of the spin gap is similar to
the $V=0$ case, i.e., it is finite for small $t_\perp$ and vanishes
for $t_\perp/t_\parallel$ larger than some critical value.
However, the charge gap is found to be nonzero for all hopping ratios
examined here
(see also Figs.~\ref{figgoptp07} and \ref{figgoptp14} below).

These data can be understood from the RG analysis of Ref.~\onlinecite{bfrg}:
the additional nearest--neighbor repulsion does not introduce a new relevant
operator, but only changes the scaling dimension of the perturbation
introduced by $U$.
This implies that small $V$ does not modify the phases found at $V=0$.
However, our data indicate that relatively small values of $V$ are
enough to drive the system to an insulating state even for
$t_\perp < t_\parallel$.

\subsection{CDW phases and charge ordering transition}
\label{sec:co}

As the nearest--neighbor repulsion $V$ is increased, we expect a
transition to a checkerboard charge--ordered state.
As this has been examined in our earlier paper, \cite{VHN99}
here we summarize the main findings:
At large $V$, an insulating CDW state with ordering wavevector
${\bf Q} = (\pi,\pi)$ occurs
for all fillings between quarter and half--filling.
At quarter--filling, the transition is second order,
i.e., the order parameter
$\eta = \lim_{L\rightarrow\infty} C({\bf Q})/\langle n\rangle^2$
vanishes continuously upon approaching a critical $V_c(U)$ from
above.
Interestingly, the transition has been found to change
from second--order to first--order at higher band filling as a function
of $U/t$; \cite{VHN99} such tricritical behavior has also been
observed in the 1D case at half--filling.\cite{Cannon91,Zhang97}
In the quarter filled CDW state, the spin correlations indicate
zigzag antiferromagnetic ordering of the spins $1 \over 2$ on the
occupied sites; at larger filling the spin correlations become
incommensurate and are gradually suppressed.

\begin{figure}
\epsfxsize=7.5 truecm
\centerline{\epsffile{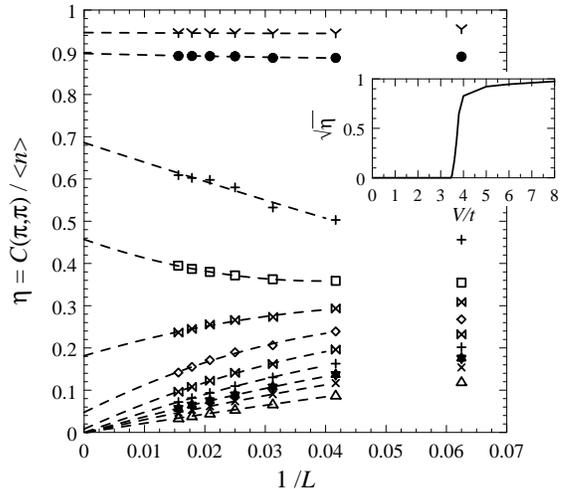}}
\caption{
  Finite--size scaling for the staggered charge
  correlation function at $U/t=4$ and isotropic hopping.
  The curves correspond to $V/t=3.0,3.2,3.3,3.4,3.5,3.6,3.7,3.8,4.0,6.0,$
  and 8.0, from bottom to top.
  The dashed lines are quadratic fits through the data points for $L\ge 24$.
  The inset shows the extrapolated values for $\sqrt{\eta}$ indicating
  a second--order transition at $V_c/t= 3.45 \pm 0.1$.
}
\label{FIG_SCALE1}
\end{figure}

We now turn to our new results and a more detailed discussion
of the quarter--filled system.
In Fig.~\ref{FIG_SCALE1}, the finite--size scaling for $\eta$
as well as the extrapolated values for $\sqrt{\eta}$,
which corresponds to the relative difference of the
sublattice occupancies in the broken--symmetry charge--ordered state,
are shown for a typical second--order transition.
(Data for first--order transitions at larger filling are shown
in Ref.~\onlinecite{VHN99}, Figs.~2 and 3.)
The inclusion of the quadratic fit term turns out to be important
near the transition.\cite{etaremark}
The $V_c(U)$ values obtained from the numerics are displayed as
phase boundary in Fig.~\ref{figpd}.
Note that $V_c$ decreases with increasing $U$ in the quarter--filled
case treated here, similar to the behavior found in 1D.\cite{Penc94}
Our results for large $U$ suggest that $V_c$ is nonzero in the
$U\rightarrow\infty$ limit as in 1D; an extrapolation based on data up to
$U/t=64$ yields $V_c(U=\infty) \approx 2 t$.
It is interesting to contrast this with {\em half filling}, for which
weak-- and strong--coupling approximations
\cite{Dongen94,Hartree} as well as numerical studies \cite{Zhang97,VHN99}
yield $V_c \approx U/z_0$ for a hypercubic lattice with $z_0$ being the number
of nearest neighbors, i.e., in the half--filled case
$V_c$ {\em increases} with $U$.

The behavior of the low--energy electronic excitations
in the vicinity of $V_c$ provides further information on the character of
the charge--order transition.
Since the energy gaps show different behavior for different values of
the hopping ratio $t_\perp/t_\parallel$, as seen in
Fig.~\ref{figgaptp}, we focus on two representative values of
the hopping anisotropy.

\begin{figure}
\epsfxsize=8 truecm
\centerline{\epsffile{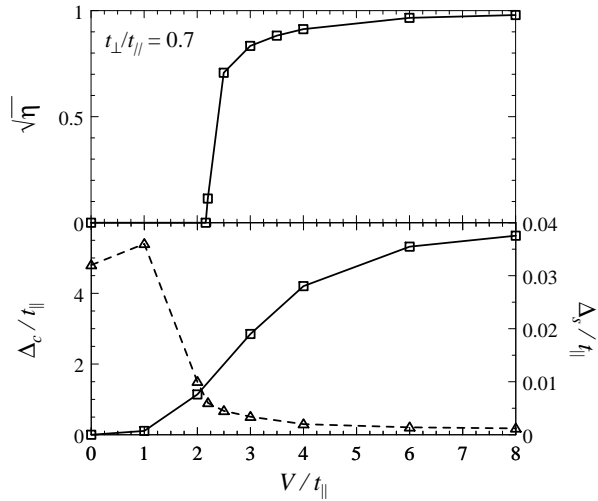}}
\caption{
  Order parameter $\sqrt{\eta}$ (upper panel) and charge (solid) and
  spin (dashed) gaps
  (lower panel) as function of $V/t_\parallel$ for $U/t_\parallel=8$ and
  $t_\perp/t_\parallel=0.7$.
  Although the spin gap decreases when entering the charge--ordered state,
  it is nonzero for all $V$.
}
\label{figgoptp07}
\end{figure}

\begin{figure}
\epsfxsize=8 truecm
\centerline{\epsffile{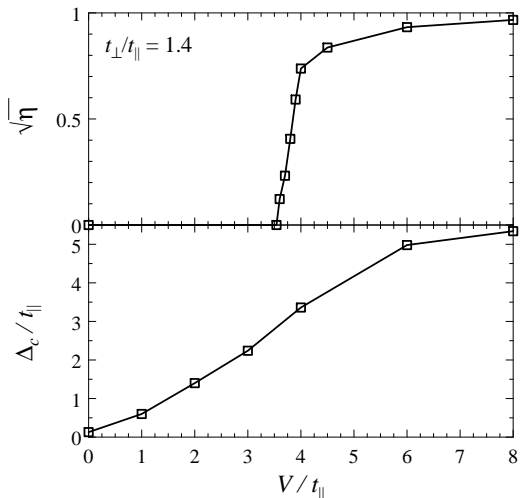}}
\caption{
  Same as Fig.~\protect\ref{figgoptp07}, but for $t_\perp/t_\parallel=1.4$.
  The lower panel now shows the charge gap only;
  the spin gap is zero within the numerical accuracy for all values of $V$.
}
\label{figgoptp14}
\end{figure}

The energy gaps as function of $V/t_\parallel$ together with the order
parameter at $U/t_\parallel=8$ are shown in Figs. \ref{figgoptp07} and
\ref{figgoptp14} for $t_\perp/t_\parallel = 0.7$ and 1.4.
The main observation is that no gap opens or closes at the
transition to the charge--ordered state.
Small $t_\perp/t_\parallel$ leads to fully gapped C0S0 phases on both
sides of the transition (\hominssg-\cdwsg\ transition),
whereas at large $t_\perp/t_\parallel$
both the homogeneous and the charge--ordered
phase have zero spin gap and nonzero charge gap
(\homins-\cdw transition).
For the former case (Fig.~\ref{figgoptp07}), the spin gap decreases
appreciably as the charge--ordered state is entered.
This indicates a change in the spin dynamics from a ``weak--coupling''
regime dominated by band effects to a ``strong--coupling'' regime
determined by the physics of a frustrated spin chain.
This behavior will be discussed in detail in Sec.~\ref{sec:strong}.

In any case, it appears that low--lying fermionic excitations do not play
a role in the critical dynamics at the charge--order transition.
This implies that this zero--temperature transition must be in the
universality class of the 1+1--dimensional Ising model.
Note, however, that inter--ladder couplings will increase the effective
dimensionality of this transition in experimental ladder systems at
low enough energies or temperatures.

\subsection{Phase separation}
\label{sec:ps}

For small $U$ and large $V$, phase separation is expected:
In the $V\to\infty$ limit, existing double occupancies are immobile
and cannot be broken up, whereas single fermions can move in
the ``unoccupied'' space.
For $U<U_{\rm PS}(V)$, the system then separates into a region
with double occupancies at every second site (i.e., checkerboard order
with two electrons per occupied site) and a region in which the other
electrons can gain kinetic energy by hopping.
For the one--dimensional system it is possible to solve the $V=\infty$
problem exactly because it maps onto non--interacting spinless fermions
moving on open chain segments. \cite{Penc94}
While the mapping to spinless fermions is similar for the ladder system,
the geometry leads to interactions among the fermions which preclude
an exact solution.
Nevertheless, the qualitative behavior should be similar to
the 1D case, i.e., for $V=\infty$ there should be a critical
$U_{\rm PS}(\infty)$ ($U_{\rm PS}(\infty)=4t$ in 1D)
below which the system phase separates.
For small $V$, the phase separation should disappear.

\begin{figure}
\epsfxsize=7 truecm
\centerline{\epsffile{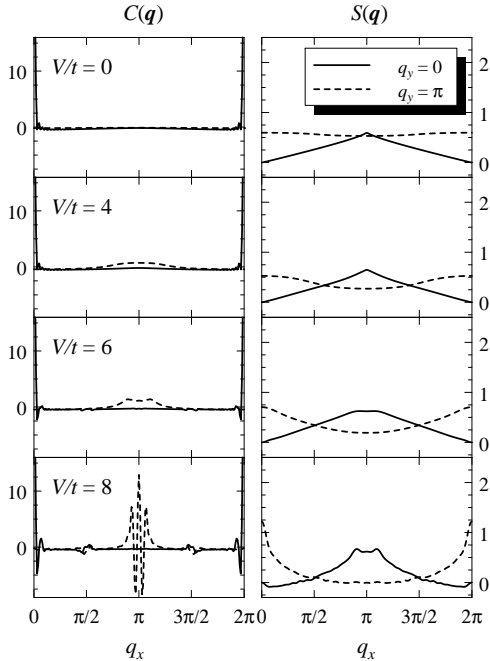}}
\caption{
 Charge and spin correlation functions for a $64 \times 2$ system with
 isotropic hopping and $U/t=2$.
 The rapid oscillations in $C({\bf q})$ at $V/t=8$ indicate that
 increasing $V$ drives the system into phase separation.
}
\label{figcorru2}
\end{figure}

The numerical results for charge and spin correlation functions
at small $U$ are shown in Fig.~\ref{figcorru2}.
Incommensurate peaks appear in both the $q_y=\pi$ channel of $C({\bf q})$
and the $q_y=0$ channel of $S({\bf q})$ as $V$ is increased.
At the largest value shown, $V/t=8$, there are strong oscillations and side
peaks in $C({\bf q})$, an indication of PS.

To examine the thermodynamic stability of the system,
we have numerically computed the
compressibility of the system which is defined as
\begin{equation}
\kappa=
\frac{4 L}{N^2} \left[
E_0(L,N+2)+E_0(L,N-2)-2E_0(L,N)
\right ] ^ {-1}
\label{kappadef}
\,.
\end{equation}
Our results clearly show the occurence of phase separation in
the large $V$, small $U$ region indicated by
(i) a diverging compressibility,
(ii) oscillating incommensurate spin and charge correlations with wave
vectors strongly dependent on the system size, and
(iii) the occurence of site charge densities greater than unity.
Note that at quarter filling no double occupancies occur even in the
perfectly charge--ordered state.
The appearance of doubly occupied sites in the phase--separated state
is clearly consistent with the phase--separation mechanism explained
above.
The criteria (i)--(iii) give consistent results and allow for a reasonably
accurate determination of the PS boundary (see Fig.~\ref{figpd}),
even though
finite--size effects in the calculation of the compressibility are
large.\cite{Penc94}

\begin{figure}
\epsfxsize=8 truecm
\centerline{\epsffile{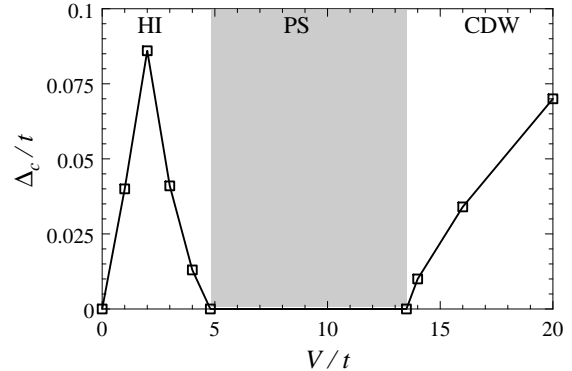}}
\caption{
  Charge gap as function of $V/t$ for $U/t=2.2$ and isotropic hopping.
  There is a homogeneous insulating phase at small $V$ and a charge--ordered
  insulating phase at large $V$; an intervening region of phase
  separation (shaded), characterized by diverging compressibility, is found
  for $4.8<V/t<13.5$.
  In the stable phases, the spin gap is zero to within the numerical
  accuracy.
}
\label{figgapu22}
\end{figure}

In contrast to that found in the 1D chain, the phase--separation
boundary has non--monotonic behavior, i.e.\ $U_{\rm PS}(V)$ shows a
maximum at around $V/t\approx 8$, $U/t \approx 2.4$, as can be seen in
Fig.~\ref{figpd}.
The described ``re--entrant'' (non--monotonic) behavior of the PS boundary
is illustrated in Fig.~\ref{figgapu22} in which
we display the charge gap
as a function of $V$ for $U/t=2.2$.
Here we find a homogeneous phase at small $V$,
a charge--ordered phase at large $V$, and a region
of phase separation in between, for $4.8<V/t<13.5$.

Another difference with the behavior of the single--chain extended
Hubbard model is that no homogeneous phase appears for large $V$
and small or intermediate values of $U$:
increasing $U$ in the PS region drives the system directly into the
charge--ordered state (Fig.~\ref{figpd}).
In contrast, in the 1D system, a homogeneous phase is present at
any $V$ and the boundaries to the charge--ordered and to the PS
phases merge (at $U/t=4$) only in the $V\to\infty$ limit. \cite{Penc94}

For the present ladder system, the behavior at the boundary between
the charge--ordered state and the PS region is quite interesting:
The charge gap appears to vanish continuously at this
boundary (Fig.~\ref{figgapu22}).
However, the charge--density wave order parameter $\sqrt{\eta}$
does not tend to zero when approaching the PS boundary from
the charge--order phase.
Moreover, the numerical results for small $U$ (in the PS region) indicate
charge--density oscillations in the spatial regions without double
occupancies.
This suggests that strong CDW correlations exist
on both sides of the PS boundary, and the transition can be
interpreted as ``continuous''.

\section{Spin dynamics in the strong--coupling limit}
\label{sec:strong}

This section focuses on the low--energy spin dynamics at large $U$ and
$V$ in the checkerboard charge--ordered state.
The charge gap in this state can be estimated to be
$\Delta_c = {\rm min}(U, 3V)$ when $t \ll U, V$,
by neglecting the kinetic energy.
In this limit, each occupied site in the charge--ordered state
carries charge $e$ and spin $1 \over 2$, and the spin states are
degenerate for $V=\infty$.
We would like to discuss the spin ordering arising from effective
exchange interactions which occur for small but finite $t/V$.
We can do this by treating $t/V$ as a perturbation, in a manner
similar to the derivation of the effective spin exchange in the
large--$U$ Hubbard model at {\em half--filling} which leads to the
mapping to an antiferromagnetic Heisenberg model.
However, the present problem is slightly more complicated because the
degeneracy is lifted in fourth order in the hopping rather than in
second order as in the half--filled Hubbard model.

The aim is to find an effective Hamiltonian for the residual spin
degrees of freedom.
It is easy to see that this model will be a frustrated
antiferromagnetic Heisenberg $J_1$--$J_2$ chain where $J_1$ and $J_2$
are a diagonal (1,1) and a horizontal (2,0) coupling between the spins in
the checkerboard ordered state.
It is well--known \cite{jjchain0,jjchain1,jjchain2,jjchain3} that this
model has a zero--temperature phase transition as a function of
$\alpha = J_2 / J_1$.
For $\alpha < \alpha_c$ the ground state is gapless with power--law
correlations.
For $\alpha > \alpha_c$ a spontaneous dimerization
occurs which leads to a spin gap and a doubly degenerate ground state.
The numerical estimate\cite{jjchain1} for $\alpha_c$ is 0.2411.
At $\alpha=1/2$ (the Majumdar--Ghosh point), the ground state has been
shown to be an exact product of nearest--neighbor
singlets.\cite{MajumdarGhosh}
Therefore, a corresponding spin--gap transition
is also possible in the charge--ordered state of the $t-U-V$ ladder
(i.e., a \cdw\ -- \cdwsg\ transition)
provided that the effective $\alpha$ can be tuned through the critical
value by changing the system parameters.

\begin{figure}
\epsfxsize=7 truecm
\centerline{\epsffile{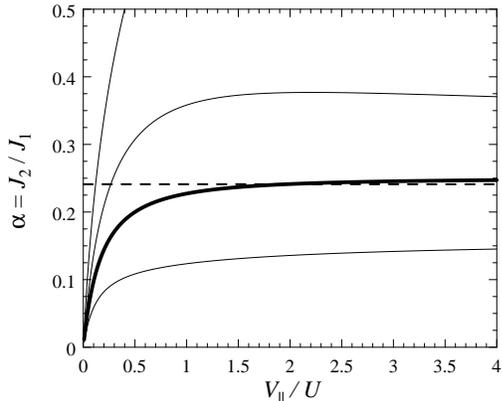}}
\caption{
The ratio of the effective coupling constants $\alpha = J_2/J_1$ calculated
from Eq. (\protect\ref{j1j2reslong}) at isotropic hopping.
The different curves
correspond to $V_\perp/V_\parallel=0$, 0.5, 1
[thick line -- see Eq. (\protect\ref{j1j2res})]
and 2 from top to bottom.
In order to obtain values for anisotropic hopping, $\alpha$ must be
multiplied by $(t_\parallel/t_{\perp})^2$.
The horizontal dashed line marks the critical value $\alpha_c=0.2411$ for the
dimerization transition of the frustrated spin chain.
}
\label{figj1j2eff}
\end{figure}

We use a recently developed method \cite{arnd} based on cumulants to
derive an effective Hamiltonian for the spin degrees of freedom in the
charge--ordered state of the quarter--filled model.
We give the derivation in the appendix, and here state only the final
result for isotropic nearest--neighbor repulsion,
$V_\parallel = V_\perp = V$:
\begin{eqnarray}
J_{1}&=&\frac{2\,t_{\perp}^{2}\, t_{\parallel}^{2}}{V^{2}}
    \left\{\frac{2}{U}+
    \frac{2}{U+2V}+\frac{1}{V}\right\}
\nonumber \\
J_{2}&=&\frac{t_{\parallel}^{4}}{V^{2}}
    \left\{\frac{1}{U}+\frac{2}{U+2V} \right\}.
\label{j1j2res}
\end{eqnarray}
The lowest--order nonzero contributions to $J_1$ and $J_2$ are
indeed of order $t^4/V^3$ and terms of order $t^4/(V^2 U)$ also appear.
For anisotropic $V$, the general expressions become more complicated
and are given in the appendix.
It turns out that the ratio $\alpha = J_2/J_1 < 1/4$
for the isotropic case $t_\parallel = t_{\perp}$ and
$V_\parallel=V_\perp$;
it approaches 1/4 for $V\gg U \gg t$, as shown in Fig.~\ref{figj1j2eff}.
The plot shows that the dimerization transition will take place at
$V/U\approx 2$.
However, this transition is hard to observe
numerically since the induced gap
is very small as we discuss below.
To access larger values of $\alpha$ a hopping anisotropy
$t_\perp/t_\parallel < 1$ is necessary.
The parameter $\alpha$ can be tuned to any value by varying the
hopping ratio.

To verify the expressions for $J_1$ and $J_2$ given above, we have studied
the behavior of the charge--ordered state in the strong--coupling limit
for different hopping anisotropies.
In order to interpret results for the spin gap, it is important to
note that $\Delta_s$ in the $J_1$--$J_2$ chain vanishes exponentially near the
critical point $\alpha_c$, leading to nonzero but very small values for
$\alpha < 0.3$.
Therefore, the spin gap calculations in the charge--ordered state require
an anisotropy in $t$ or $V$ in order to reach $\alpha$ values
significantly larger than 0.3.
Furthermore, they are feasible only in a window of intermediate values of
$U/t, V/t$:
overly small values do not lead to a charge--ordered state,
whereas overly large values of $V$ lead to an unobservably small spin
gap (of order $J \sim t^4/V^3$).

The nature of the magnetic ground state can also be probed using static
spin correlation functions.
For the $J_1$--$J_2$ chain it is known \cite{jjchain2,jjchain3} that the
static structure factor $S(q)$ is peaked
at $q = \pi$ for $\alpha<1/2$.
For $\alpha>1/2$, the peak position shifts to smaller $q$ as
$\alpha$ is increased, approaching $\pi/2$, the value for two uncoupled
chains with a doubled lattice constant, as $\alpha$ becomes large.

\begin{figure}
\epsfxsize=7 truecm
\centerline{\epsffile{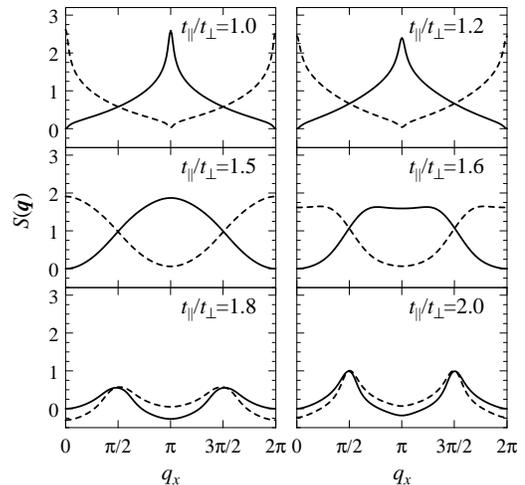}}
\caption{
Spin correlation functions $S({\bf q})$ in the charge--ordered state of
the quarter--filled ladder.
As in Fig. \ref{figcorru2}, $q_y = 0,\pi$ data are plotted using full and
dashed lines, respectively.
The parameters are $L=64$ and $U=V=8 t_\perp$,
the different values of the hopping ratio $t_\parallel/t_\perp$
correspond to effective exchange constants
[from (\protect\ref{j1j2res})] with
$\alpha = J_2/J_1 = 0.227$, 0.327, 0.511, 0.582, 0.736, and 0.909.
}
\label{figj1j2sq}
\end{figure}

\begin{figure}
\epsfxsize=9.3 truecm
\centerline{\epsffile{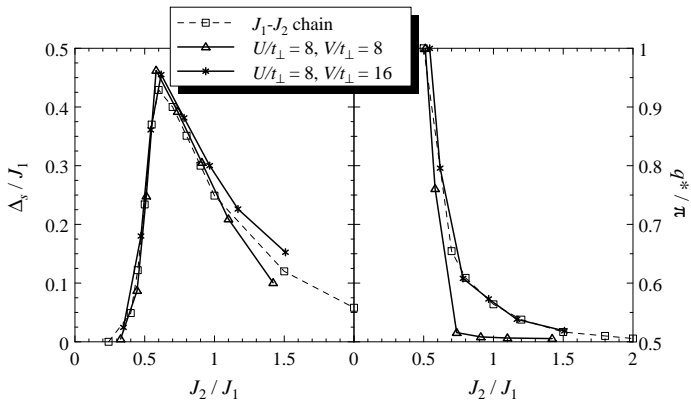}}
\caption{
Spin gap (left) and peak position in the spin structure factor (right)
for the charge--ordered state of the quarter--filled $t-U-V$ ladder.
The different curves are obtained by varying the hopping ratio
$t_\parallel/t_\perp$ at fixed values of $U/t_\perp$ and $V/t_\perp$.
The horizontal axis shows the ratio of the effective exchange
constants $J_1$ and $J_2$ obtained from the strong--coupling
expressions (\protect\ref{j1j2res}).
Data marked "$J_1$--$J_2$ chain" are taken from
Ref.\ \protect\onlinecite{jjchain3}.
}
\label{figj1j2dim}
\end{figure}

The spin correlations at $U=V=8 t_\perp$ and various hopping ratios
obtained from the DMRG calculations are shown in
Fig.~\ref{figj1j2sq}.
(Note that we use $t_\perp$ rather than $t_\parallel$ as the
energy reference in this section.)
A shift of the maximum from $q_x=\pi$ to $q_x=\pi/2$ (at $q_y=0$) with
increasing $t_\parallel/t_\perp$ is clearly visible.
To compare quantitatively with the strong--coupling picture of the
frustrated spin chain,
we show in Fig.~\ref{figj1j2dim} the results for the spin gap $\Delta_s$
and for the peak position $q^{*}$ in the spin structure factor
(see Fig.~\ref{figj1j2sq}) for different parameter sets in the
charge--ordered phase.
Data for the frustrated spin chain from Ref.\ \onlinecite{jjchain3} are
also shown for comparison.
Note that the data are plotted as a function of the ratio $J_2/J_1$ with
the values of these {\em effective} couplings taken from the
strong--coupling expressions (\ref{j1j2res}).
The spin gap value follows the strong coupling prediction closely even for
intermediate values of $U/t$ and $V/t$.
The peak position also shows the expected behavior, i.e., it deviates from
$\pi$ when the effective $J_2/J_1$ exceeds a certain value.
For large $U$ and $V$, the agreement with the results from the
frustrated spin chain is nearly perfect, clearly indicating that
the spin dynamics in the strong--coupling charge--ordered state is
correctly described by the $J_1$--$J_2$ spin chain.
For smaller values of $U$ and $V$, there are slight deviations in the
peak position from the spin chain data: the region of
incommensurate spin order becomes narrower with decreasing
interaction.
This might be expected because there is no incommensurability at
half--filling in the non--interacting limit.
A similar behavior for $S(q)$ has been found for the half--filled
Hubbard chain with next--nearest--neighbor hopping \cite{t1t2hub}
which can also be mapped onto an effective frustrated spin chain in
the large--$U$ limit.

Now we turn to a discussion of the special case of isotropic hopping
for which the $V=0$ weak coupling system is near the one--band to
two--band transition.
The numerical results obtained by DMRG indicate a zero spin gap and
nonzero charge gap for any finite $V$ (outside the phase--separation
region).
This is in disagreement with the strong--coupling analysis presented
above which predicts a \cdwsg\ phase at large $V/U$ due to spontaneous
dimerization.
Since $\alpha=J_2/J_1$ is close
to $\alpha_c$, however, the spin gap would be very small and therefore
hard to observe using numerical methods.
By assuming that the strong--coupling picture is also valid in the
intermediate--coupling regime, we can locate the \cdw\ -- \cdwsg\ boundary
as shown in Fig.\ \ref{figpd}.
Since it is not possible to deduce the behavior of the spin gap close
to the charge--order transition from the current numerical results,
we cannot decide whether the charge--order transition line and the
spin--gap transition line meet for the case of isotropic hopping.
Additional numerical approaches (e.g., based on level--crossing methods)
could be used to check the spin--gap scenario and to determine the
precise location of the boundary of the spin--gap phase.

It is worth pointing out that although a spin gap is present
in both the homogeneous and the charge--ordered phases at small
$t_\perp/t_\parallel$, the mechanisms for the spin gap opening
appear to be quite different:
The strong--coupling \cdw\ -- \cdwsg\ transition involves spontaneous
dimerization in a spin model and is described by a sine--Gordon theory,
whereas the weak--coupling case at $V=0$ is more complicated due to the
presence of low--lying charge modes (see Ref.\ \onlinecite{bfrg} for
a discussion on the RG for $V=0$), however, not much is known about the
$V>0$ case.


\section{Conclusions}
\label{sec:concl}

In summary, we have studied the phase diagram of the extended
Hubbard model on a quarter--filled two--leg ladder.
At very small $V$, the system behaves as in the $V=0$ case \cite{bfrg},
while slightly larger values of $V$ lead to an insulating state with
either zero (\homins) or nonzero (\hominssg) spin gap depending on the
hopping anisotropy.
For $U$ and $V$ both strong, the ground state shows zigzag charge
order.
In this phase each occupied site carries spin $1 \over 2$ and the
residual kinetic energy leads to effective antiferromagnetic exchange
interactions between the spins.
We have rigorously established the mapping of the spin degrees of
freedom to a frustrated spin--$\frac{1}{2}$ chain in
the strong coupling limit, $U, V \gg t$.
This effective spin chain can be either in the gapless regime (\cdw)
with algebraic spin correlations or in the spontaneously dimerized
regime (\cdwsg) with gapped spin excitations.
For $t_\perp < t_\parallel$, the spin gap in
the charge--ordered state could be numerically observed.
Its magnitude is in good agreement with the results for a
corresponding $J_1$--$J_2$ spin chain down to $U/t=4$.
The dimerization of the effective spin chain can be interpreted
as bond--order wave \cite{mazumdar} in the original Hubbard model,
so the system has an insulating \cdwsg\ ground state with coexisting
bond--order and charge--density waves.
Finally, at small values of $U$ and moderate to large values of $V$,
the system phase separates into a phase of immobile double
occupancies on every second site and a phase of mobile single
electrons.

We have identified a purely electronic mechanism for the opening of a
spin gap in a quarter--filled CDW system on a ladder based on the
physics of a frustrated spin chain.
However, we note here that the spin--gap physics discussed in
Sec.~\ref{sec:strong} probably cannot be realized in NaV$_2$O$_5$
since this material has $t_\perp \approx 2 t_\parallel$ (see
Refs.~\onlinecite{Smol98,khomskii}) leading to $J_1 \gg J_2$ for the
effective spin chain.
It is likely that the spin gap opening in NaV$_2$O$_5$ is driven by
the interplay of charge ordering and phonons, as suggested in
Ref.\ \onlinecite{khomskii}.
Other effects which are important for the spin dynamics in
NaV$_2$O$_5$ are hopping processes between the ladders which may lead
to quite large exchange terms across the ladders. \cite{thal2}

We have found no evidence for ``exotic'' phases in the examined ladder
system like the ones present in the single--chain
model.\cite{Penc94,Clay99}
There appears to be no metallic phase in the
quarter--filled model except for the one at very small $V$
and $t_\perp < t_\parallel$ (Fig.~\ref{figpdtr}).
This phase has been
discussed \cite{bfrg} in the context of the ``bare'' Hubbard model
on the ladder.

Any effects of interladder couplings have been neglected in the present
treatment, as well as the interplay of electron and lattice effects
which is known to lead to further interesting ordering effects; \cite{riera}
these should be investigated in the future.
Also, a more detailed study of the spin dynamics in a partially
charge--ordered state ($V<\infty$) could be performed.

\acknowledgments

The authors thank D.~Baeriswyl, R.~Bulla, P.~G.~J.~van Dongen,
R.~E.~Hetzel, and A.~P.~Kampf for useful conversations.
M.V. acknowledges support by the DFG (VO 794/1--1) and by US NSF Grant No
DMR 96--23181, R.M.N.\ has been supported by
the Swiss National Foundation under Grant No 20--53800.98.
The calculations were performed on the Origin 2000 at the Technical University
Dresden.


\appendix
\section{Effective Hamiltonian for the charge--ordered state}

This appendix provides the derivation of the effective exchange
Hamiltonian for the quarter--filled ladder in the strongly
charge--ordered regime, $(V,U) \gg t$.
The charge degrees of freedom are projected out, i.e., the effective
Hamiltonian ${\cal H}_{\rm eff}$ acts in a Hilbert space where every
second site is singly occupied.
This is analogous to the derivation of the Heisenberg model as
large--$U$ limit of the half--filled one--band Hubbard model.
The present problem maps onto a frustrated $J_1$--$J_2$ spin chain.
The effective exchange arises from fourth--order hopping processes
which makes the problem more complicated than the half--filled Hubbard
model for which the lowest non--trivial contributions arise at second
order in $t$.

We apply a recently developed cumulant method \cite{arnd} to derive
${\cal H}_{\rm eff}$.
It is useful to split ${\cal H} = {\cal H}_0+{\cal H}_1$
where ${\cal H}_0$ contains the dominating interaction terms and
${\cal H}_1$ the perturbation caused by hopping.
We start with broken translational symmetry from the outset and
define a projection operator ${\cal P}$ which projects onto the low--energy
space where the charges show perfect checkerboard charge order, i.e.,
$\langle n_i \rangle = [1+\exp(i {\bf Q R}_i)]/2$ with
${\bf Q} = (\pi,\pi)$.
This order defines two sublattices which we will denote as $A$ and $B$
for occupied and unoccupied, respectively.
Transitions between states within the ${\cal P}$ space are only possible with
four or more hopping processes; the fourth order processes only involve
intermediate states outside the ${\cal P}$ space.
The fourth--order Hamiltonian can be obtained by fourth--order perturbation
theory and is given by \cite{arnd}
\begin{eqnarray}
  {\cal H}_{\rm eff}&=&-{\cal P}{\cal H}_{1}{\cal Q}\frac{1}{{\cal H}_{0}}
    {\cal Q}{\cal H}_{1}{\cal Q}\frac{1}{{\cal H}_{0}}{\cal Q}{\cal H}_{1}
    {\cal Q}\frac{1}{{\cal H}_{0}}{\cal Q}{\cal H}_{1}{\cal P},
\label{effHami}
\end{eqnarray}
where ${\cal Q} \equiv 1-{\cal P}$.

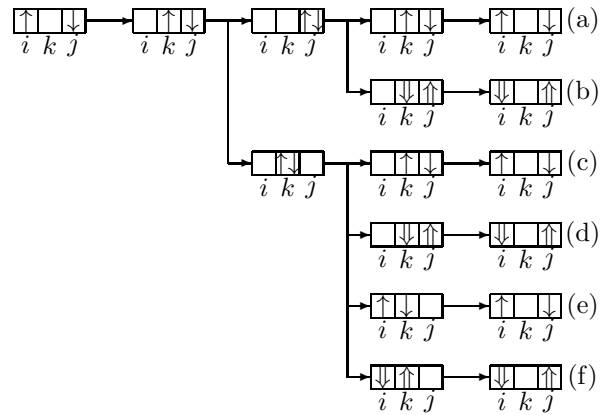
\begin{figure}
  \begin{center}
    \setlength{\unitlength}{1.8pt}
\begin{picture}(132,90)
  \multiput(5,85)(25,0){5}{\line(1,0){15}}
  \multiput(5,80)(25,0){5}{\line(1,0){15}}
  \multiput(80,70)(25,0){2}{\line(1,0){15}}
  \multiput(80,65)(25,0){2}{\line(1,0){15}}
  \multiput(55,55)(25,0){3}{\line(1,0){15}}
  \multiput(55,50)(25,0){3}{\line(1,0){15}}
  \multiput(80,40)(25,0){2}{\line(1,0){15}}
  \multiput(80,35)(25,0){2}{\line(1,0){15}}
  \multiput(80,25)(25,0){2}{\line(1,0){15}}
  \multiput(80,20)(25,0){2}{\line(1,0){15}}
  \multiput(80,10)(25,0){2}{\line(1,0){15}}
  \multiput(80,5)(25,0){2}{\line(1,0){15}}

  \multiput(5,80)(25,0){5}{\line(0,1){5}}
  \multiput(10,80)(25,0){5}{\line(0,1){5}}
  \multiput(15,80)(25,0){5}{\line(0,1){5}}
  \multiput(20,80)(25,0){5}{\line(0,1){5}}
  \multiput(80,65)(25,0){2}{\line(0,1){5}}
  \multiput(85,65)(25,0){2}{\line(0,1){5}}
  \multiput(90,65)(25,0){2}{\line(0,1){5}}
  \multiput(95,65)(25,0){2}{\line(0,1){5}}
  \multiput(55,50)(25,0){3}{\line(0,1){5}}
  \multiput(60,50)(25,0){3}{\line(0,1){5}}
  \multiput(65,50)(25,0){3}{\line(0,1){5}}
  \multiput(70,50)(25,0){3}{\line(0,1){5}}
  \multiput(80,35)(25,0){2}{\line(0,1){5}}
  \multiput(85,35)(25,0){2}{\line(0,1){5}}
  \multiput(90,35)(25,0){2}{\line(0,1){5}}
  \multiput(95,35)(25,0){2}{\line(0,1){5}}
  \multiput(80,20)(25,0){2}{\line(0,1){5}}
  \multiput(85,20)(25,0){2}{\line(0,1){5}}
  \multiput(90,20)(25,0){2}{\line(0,1){5}}
  \multiput(95,20)(25,0){2}{\line(0,1){5}}
  \multiput(80,5)(25,0){2}{\line(0,1){5}}
  \multiput(85,5)(25,0){2}{\line(0,1){5}}
  \multiput(90,5)(25,0){2}{\line(0,1){5}}
  \multiput(95,5)(25,0){2}{\line(0,1){5}}

  \multiput(20,82.5)(25,0){3}{\vector(1,0){10}}
  \multiput(95,7.5)(0,15){6}{\vector(1,0){10}}
  \put(70,52.5){\vector(1,0){10}}

  \put(50,82.5){\line(0,-1){30}}
  \put(75,82.5){\line(0,-1){15}}
  \put(75,52.5){\line(0,-1){45}}

  \put(50,52.5){\vector(1,0){5}}
  \put(75,67.5){\vector(1,0){5}}
  \multiput(75,7.5)(0,15){3}{\vector(1,0){5}}

  \multiput(5,75)(25,0){5}{\makebox(5,5){$i$}}
  \multiput(10,75)(25,0){5}{\makebox(5,5){$k$}}
  \multiput(15,75)(25,0){5}{\makebox(5,5){$j$}}
  \multiput(80,60)(25,0){2}{\makebox(5,5){$i$}}
  \multiput(85,60)(25,0){2}{\makebox(5,5){$k$}}
  \multiput(90,60)(25,0){2}{\makebox(5,5){$j$}}
  \multiput(55,45)(25,0){3}{\makebox(5,5){$i$}}
  \multiput(60,45)(25,0){3}{\makebox(5,5){$k$}}
  \multiput(65,45)(25,0){3}{\makebox(5,5){$j$}}
  \multiput(80,30)(25,0){2}{\makebox(5,5){$i$}}
  \multiput(85,30)(25,0){2}{\makebox(5,5){$k$}}
  \multiput(90,30)(25,0){2}{\makebox(5,5){$j$}}
  \multiput(80,15)(25,0){2}{\makebox(5,5){$i$}}
  \multiput(85,15)(25,0){2}{\makebox(5,5){$k$}}
  \multiput(90,15)(25,0){2}{\makebox(5,5){$j$}}
  \multiput(80,0)(25,0){2}{\makebox(5,5){$i$}}
  \multiput(85,0)(25,0){2}{\makebox(5,5){$k$}}
  \multiput(90,0)(25,0){2}{\makebox(5,5){$j$}}

  \put(5,80){\makebox(5,5){$\uparrow$}}
  \put(15,80){\makebox(5,5){$\downarrow$}}
  \put(35,80){\makebox(5,5){$\uparrow$}}
  \put(40,80){\makebox(5,5){$\downarrow$}}
  \put(65,80){\makebox(5,5){$\uparrow\downarrow$}}
  \put(85,80){\makebox(5,5){$\uparrow$}}
  \put(90,80){\makebox(5,5){$\downarrow$}}
  \put(105,80){\makebox(5,5){$\uparrow$}}
  \put(115,80){\makebox(5,5){$\downarrow$}}

  \put(85,65){\makebox(5,5){$\Downarrow$}}
  \put(90,65){\makebox(5,5){$\Uparrow$}}
  \put(105,65){\makebox(5,5){$\Downarrow$}}
  \put(115,65){\makebox(5,5){$\Uparrow$}}

  \put(60,50){\makebox(5,5){$\uparrow\downarrow$}}
  \put(85,50){\makebox(5,5){$\uparrow$}}
  \put(90,50){\makebox(5,5){$\downarrow$}}
  \put(105,50){\makebox(5,5){$\uparrow$}}
  \put(115,50){\makebox(5,5){$\downarrow$}}

  \put(85,35){\makebox(5,5){$\Downarrow$}}
  \put(90,35){\makebox(5,5){$\Uparrow$}}
  \put(105,35){\makebox(5,5){$\Downarrow$}}
  \put(115,35){\makebox(5,5){$\Uparrow$}}

  \put(80,20){\makebox(5,5){$\uparrow$}}
  \put(85,20){\makebox(5,5){$\downarrow$}}
  \put(105,20){\makebox(5,5){$\uparrow$}}
  \put(115,20){\makebox(5,5){$\downarrow$}}

  \put(80,5){\makebox(5,5){$\Downarrow$}}
  \put(85,5){\makebox(5,5){$\Uparrow$}}
  \put(105,5){\makebox(5,5){$\Downarrow$}}
  \put(115,5){\makebox(5,5){$\Uparrow$}}

  \put(122,80){\makebox(5,5){(a)}}
  \put(122,65){\makebox(5,5){(b)}}
  \put(122,50){\makebox(5,5){(c)}}
  \put(122,35){\makebox(5,5){(d)}}
  \put(122,20){\makebox(5,5){(e)}}
  \put(122,5){\makebox(5,5){(f)}}

\end{picture}
  \end{center}
  \caption{
     Processes contributing to $J_2$.
     Here, $i,j\in A$ are the originally occupied lattice sites whereas
     $k\in B$ is the intermediate site. $\Uparrow$ und $\Downarrow$ denote
     spins being reversed with respect to the initial configuration.
     }
\label{j2proc}
\end{figure}

This expression can be easily used to identify the processes
contributing to the effective exchange.
We first discuss the fourth--order processes leading to $J_2$, i.e.,
the processes coupling two spins located on the same leg of the
ladder.
They involve exactly one site in between the two originally occupied
sites.
Therefore, any fourth order hopping process must involve a
temporary double occupancy.
A suitable classification of the possible processes is shown in
Fig.\ \ref{j2proc}.

Any process involves three transition states giving rise to the energy
denominator of the final expression for $J_2$.
It is easy to see that (a) and (b) have transition states with
energies $V_\parallel+V_\perp$, $U$, and $V_\parallel+V_\perp$ (the
$V_\perp$ arises from the nearby occupied site on the second leg)
whereas (c)--(f) have transition state energies $V_\parallel+V_\perp$,
$U+2V_\perp$, and $V_\parallel+V_\perp$.
The sum of all processes has the form
$J_2 ( n_{i,\sigma}n_{j,-\sigma}+
c_{i,\sigma}^{\dagger}c_{i,-\sigma}c_{j,\sigma}
c_{j,-\sigma}^{\dagger})$.
Examination of the signs shows that the resulting exchange is
antiferromagnetic, $J_2>0$.
For the $J_1$ processes there are two intermediate empty sites in
between the two occupied sites
Therefore, processes involving one or both of these two
sites are possible.
Particularly interesting are the circular hopping processes
contributing to the diagonal exchange $J_1$ in which both intermediate
sites are involved and {\em no} temporary double occupancy occurs.
These processes include four sites and the two electrons under
consideration are never on the same site.
Nevertheless, these processes lead to an effective spin--spin
interaction ${\bf S}_1\cdot{\bf S}_2$ (and not only a constant energy
shift):
For parallel spins the process reproduces the original state, whereas
antiparallel spins are {\em always} exchanged.
This leads to the exchange form
$S_1^-S_2^+ + S_1^+S_2^- + 2 (S_1^z S_2^z + 1/2)$.

For isotropic hopping and interaction, these considerations can be summarized
to $J_1 = 2 J_2 + J_{\rm circ}$, where $J_{\rm circ}$ arises from the circular
hopping processes;
in the general anisotropic case the $J_1$ processes will have energy
denominators different from the ones quoted above.
Collecting all terms leads to the following (general) result for the exchange
couplings:
\begin{eqnarray}
J_{1}&=& 4\,t_{\perp}^{2}\, t_{\parallel}^{2}
    \left\{ \left( \frac{1}{U} + \frac{1}{U+2V_\parallel} \right) \,
    \left(\frac{1}{(2V_\parallel)^2} +
       \frac{1}{V_\parallel(V_\perp+V_\parallel)}\, + \right.\right.
\nonumber\\
    &&~~~~~~~~~~~~\left.\left.
    + \frac{1}{(V_\parallel+V_\perp)^2}\right )
    + \frac{1}{V_\parallel^2(V_\perp+V_\parallel)}
    \right\}
\nonumber \\
J_{2}&=&\frac{4\,t_{\parallel}^{4}}{(V_\parallel+V_\perp)^{2}}
    \left\{\frac{1}{U}+\frac{2}{U+2V_\perp} \right\}.
\label{j1j2reslong}
\end{eqnarray}
For isotropic repulsion, $V_\parallel= V_\perp$, these expressions reduce to
Eq. (\ref{j1j2res}) quoted in the body of the paper.



\end{multicols}


\begin{thebibliography}{99}

\bibitem[*]{addr} New permanent address: Theoretische Physik III,
        Elektronische Korrelationen und Magnetismus, Universit\"{a}t
        Augsburg, D--86135 Augsburg, Germany.

\bibitem{ohama} T.~Ohama, H.~Yasuoka, M.~Isobe, and Y.~Ueda, \prb
        {\bf 59}, 3299 (1999).

\bibitem{isobe} M.~Isobe and Y.~Ueda, J. Phys. Soc. Jpn. {\bf 65}, 1178 (1996).

\bibitem{neutron} Y.~Fujii, H.~Nakao, T.~Yosihama, M.~Nishi,
        K.~Nakajima, K.~Kakurai, M.~Isobe, Y.~Ueda, and H.~Sawa,
        J. Phys. Soc. Jpn. {\bf 66}, 326 (1997).

\bibitem{Smol98} H.~Smolinski, C.~Gros, W.~Weber, U.~Peuchert,
        G.~Roth, M.~Weiden,
        and C.~Geibel, \prl {\bf 80}, 5164 (1998).

\bibitem{Seo98} H.~Seo and H.~Fukuyama, J. Phys. Soc. Jpn. {\bf 67},
        2602 (1998).

\bibitem{dagottoreview} E.~Dagotto and T.~M.~Rice, Science {\bf 271},
                        618 (1996).


\bibitem{ohama2} T.~Ohama, A.~Goto, T.~Shimizu, E.~Ninomiya, H.~Sawa,
        M.~Isobe, and Y.~Ueda, cond--mat/0003141.

\bibitem{nakao} H.~Nakao, K.~Ohwada, N.~Takesue, Y.~Fujii, M.~Isobe,
        Y.~Ueda, M.~v.~Zimmermann,
        J.~P.~Hill, D.~Gibbs, J.~C.~Woicik, I.~Koyama, and
        Y.~Murakami, cond--mat/0003129.

\bibitem{grenier00} B.~Grenier, O.~Cepas, L.~P.~Regnault,
        J.~E.~Lorenzo, T.~Ziman, J.~P.~Boucher, A.~Hiess,
        T.~Chatterji, J.~Jegoudez, and A.~Revcolevschi, cond--mat/0007025.


\bibitem{gros} C.~Gros and R.~Valenti, \prl {\bf 82}, 976 (1999).


\bibitem{thal2} P.~Thalmeier and A.~N.~Yaresko, Eur. Phys. J. B
        {\bf 14}, 495 (2000).

\bibitem{smaalen} S.~van Smaalen and J.~L\"{u}decke,
        Europhys. Lett. {\bf 49},       250 (2000).


\bibitem{steglich1} M.~K\"{o}ppen, D.~Pankert, R.~Hauptmann, M.~Lang,
        M.~Weiden, C.~Geibel, and F.~Steglich, \prb {\bf 57}, 8466
        (1998).

\bibitem{thal1} P.~Thalmeier and P.~Fulde, Europhys. Lett. {\bf 44},
        242 (1998).


\bibitem{boer00} J.~L.~de Boer, A.~Meetsma, J.~Baas, and T.~T.~M.~Palstra,
       \prl {\bf 84}, 3962 (2000).

\bibitem{gros00} C.~Gros, R.~Valenti, J.~V.~Alvarez, K.~Hamacher,
        and W.~Wenzel, cond--mat/0004404.


\bibitem{hubbard78} J.~Hubbard, \prb {\bf 17}, 494 (1978).


\bibitem{Mila93} F.~Mila and X.~Zotos, Europhys. Lett. {\bf 24}, 133 (1993).

\bibitem{Penc94} K.~Penc and F.~Mila, Phys. Rev. B {\bf 49}, 9670 (1994).

\bibitem{Cannon91} J.~W.~Cannon, R.~T.~Scalettar, and E.~Fradkin,
        Phys. Rev. B {\bf 44}, 5995 (1991).

\bibitem{hirsch} J. E. Hirsch, \prl {\bf 53}, 2327 (1984);
        \prb {\bf 31}, 6022 (1985).

\bibitem{Zhang97} G~ P.~Zhang, Phys. Rev. B {\bf 56}, 9189 (1997).

\bibitem{tricrit} M.~Nakamura, J. Phys. Soc. Jpn. {\bf 68}, 3123 (1999),
        \prb {\bf 61}, 16377 (2000); cond--mat/0003419.

\bibitem{Lin95} H.~Q.~Lin, E.~R.~Gagliano, D.~K.~Campbell,
        E.~H.~Fradkin, and J.~E.~Gubernatis in
        {\it The Hubbard Model, its Physics and Mathematical Physics,}
        edited by D.~Baeriswyl {\it et al.}, NATO ASI Series (Plenum,
        New York, 1995).

\bibitem{Clay99} R. T. Clay, A. W. Sandvik, and D. K. Campbell,
        Phys. Rev. B {\bf 59}, 4665 (1999).

\bibitem{Zhang89} Y. Zhang and J. Callaway, Phys. Rev. B {\bf 39}, 9397 (1989).

\bibitem{Chatto97} B.~Chattopadhyay and D.~M.~Gaitonde,
        \prb {\bf 55}, 15364 (1997).

\bibitem{Pietig99} R. Pietig, R. Bulla, and S. Blawid, \prl {\bf 82},
        4046 (1999).

\bibitem{Dongen94}P.~G.~J.~van Dongen, Phys. Rev. B {\bf 49}, 7904 (1994);
        Phys. Rev. B {\bf 50}, 14016 (1994).

\bibitem{Dongen96}P.~G.~J.~van Dongen, Phys. Rev. B {\bf 54}, 1584 (1996).

\bibitem{VHN99} M.~Vojta, R.~E.~Hetzel, and R.~M.~Noack, \prb {\bf 60},
        R8417 (1999).


\bibitem{bfrg} L.~Balents and M.~P.~A.~Fisher, \prb {\bf 53}, 12133 (1996).

\bibitem{noacktwochains} R.~M.~Noack, S.~R.~White, and D.~J.~Scalapino,
        Physica C {\bf 270}, 281 (1996).


\bibitem{DMRG} S.~R.~White, \prl {\bf 69}, 2863 (1992),
    \prb {\bf 48}, 10345 (1993).

\bibitem{etaremark}
  The present $V_c$ values are more accurate and slightly larger than the
  ones reported in Ref.~\onlinecite{VHN99}; this is due to more data points
  and the additional quadratic term for the $1/L$ finite--size extrapolation.

\bibitem{Hartree} D.~Cabib and E.~Callen, \prb {\bf 12}, 5249 (1971);
        R.~A.~Bari, \prb {\bf 3}, 2662 (1971).


\bibitem{jjchain0} F.~D.~M.~Haldane, \prb {\bf 25}, 4925 (1982),
        I.~Affleck, D.~Gepner, H.~J.~Schulz, and T.~Ziman, J. Phys. A
        {\bf 22}, 511 (1989).

\bibitem{jjchain1} S.~Eggert and I.~Affleck, \prb {\bf 46}, 10866 (1992),
        S.~Eggert, \prb {\bf 54}, 9612 (1996).

\bibitem{jjchain2} R.~Chitra, S.~Pati, H.~R.~Krishnamurthy, D.~Sen, and
        S.~Ramasesha, \prb {\bf 52}, 6581 (1995).

\bibitem{jjchain3} S.~R.~White and I.~Affleck, \prb {\bf 54}, 9862 (1996).

\bibitem{MajumdarGhosh} C. K. Majumdar and D. K. Ghosh,
        J. Math. Phys. {\bf 10}, 1388 (1969).


\bibitem{arnd} A.~H\"{u}bsch, M.~Vojta, and K.~W.~Becker, J. Phys. Cond. Matter
        {\bf 11}, 8523 (1999).


\bibitem{t1t2hub} S.~Daul and R.~M.~Noack, \prb {\bf 61}, 1646 (2000).


\bibitem{mazumdar} S.~Mazumdar, S.~Ramasesha, R.~T.~Clay, D.~K.~Campbell,
        \prl {\bf 82}, 1522 (1999);
        S.~Mazumdar, R.~T.~Clay, and D.~K.~Campbell, cond--mat/9910164;
        cond--mat/0003200.

\bibitem{khomskii} M. V. Mostovoy and D. I. Khomskii,
        Solid State Commun. {\bf 113}, 159 (2000).

\bibitem{riera} J.~Riera and D.~Poilblanc, \prb {\bf 59}, 2667 (1999).


\end{thebibliography}
\end{document}